\documentclass[fleqn,twoside]{article}
\usepackage{espcrc2}

\usepackage{graphicx}

\newcommand{\kms}{km~s$^{-1}\,$}
\newcommand{\ms}{m~s$^{-1}\,$}

\newcommand{\cmm}{cm$^{-3}\,$}

\newcommand{\dmm}{$\Delta\mu/\mu\,$}

\newcommand{\AmS}{{\protect\the\textfont2A\kern-.1667em\lower.5ex\hbox{M}\kern-.125emS}}

\newcommand{\ndhpb}{N$_2$H$^+$ }
\newcommand{\nddp}{N$_2$D$^+$}
\newcommand{\nddpb}{N$_2$D$^+$ }

\newcommand{\juz}{(J:1--0)}
\newcommand{\juzb}{(J:1--0) }
\newcommand{\jdu}{(J:2--1)}

\newcommand{\jtd}{(J:3--2)}

\title{Stringent bounds to spatial variations of the
electron-to-proton mass ratio in the Milky Way
}
 
\author{P.~Molaro\address[INAF-OAT]{INAF-Osservatorio Astronomico di Trieste, \\
Via G. B. Tiepolo 11, 34143 I, Trieste, Italy},
S. A. Levshakov\address[Ioffe]{Ioffe Physical-Technical Institute, \\ Politekhnicheskaya Str. 26,
              194021 St. Petersburg, Russia}, and %
M. G. Kozlov\address[]{Petersburg Nuclear Physics Institute, \\ Gatchina, 188300, Russia}
       }
       
\begin{document}

\begin{abstract}
The ammonia method, recently proposed by Flambaum and Kozlov (2007) to probe 
  variations of the
electron-to-proton mass ratio, $\mu = m_{\rm e}/m_{\rm p}$, is applied for the first time  to   
dense prestellar molecular
clouds in the Milky Way, allowing to 
  test $\Delta\mu/\mu$ 
at different galactocentric distances.
High quality radio-astronomical observations  are used to check the presence of possible  relative radial
velocity offsets
  between the inversion transition of 
NH$_3$ $(J,K) = (1,1)$, 
  and the CCS $J_N = 2_1-1_0$ and N$_2$H$^+$ $J = 1-0$ rotational transitions. 
Carefully selected  sample of 21  NH$_3$/CCS pairs observed in the
Perseus molecular cloud  provide
  the offset
$\Delta V_{{\rm CCS-NH}_3}$$= 36\pm7_{\rm stat}\pm13.5_{\rm sys}$ \ms\ .
A similar offset  of   $\Delta V = 40.8\pm 12.9_{\rm stat} $ \ms\  between NH$_3$ $(J,K) = (1,1)$ 
and N$_2$H$^+$ $J = 1-0$ has been found  in 
  an isolated
dense core L183 by Pagani et al. (2009).

Overall these observations  provide  a safe bound  of  a maximum offset between  ammonia and the other molecules  
at the level of  $\Delta V \le 100  $ \ms.
  Being interpreted in terms of $\Delta\mu/\mu$, this bound corresponds 
to   \dmm $ \le 1\times 10^{-7}$,  
  which is  an order of magnitude 
more sensitive than available extragalactic  constraints. 
Taken at face value the measured $\Delta V$ shows  positive  shifts between
the line centers of NH$_3$ and these two other molecules and suggest a real offset, 
which  would imply  a \dmm $\sim 4\times10^{-8}$.  
If \dmm\ follows the gradient of the local gravitational potential,   then 
the obtained results  are  in conflict  with laboratory atomic clock experiments in the solar system 
by $\sim 5$ orders of magnitude, thus requiring    
  a chameleon-type scalar field model. 
 New  measurements involving  other molecules and a wider range of objects 
along with verification of molecular rest frequencies 
are currently planned   to confirm these first indications.  

 \vspace{1pc}
\end{abstract}

\maketitle

\section{Introduction}

The  late  acceleration in the universal expansion reveals a  negative-pressure component of the bulk
of energy density,  `dark energy'. This can be  a cosmological constant or 
 a  dynamically evolving scalar field $\varphi$, `quintessence' , which 
 does not have a   cosmic coincidence problem.
 However,  in quintessence the coupling of  scalar field  to ordinary matter leads 
unavoidably to long-range forces with a variability of the physical
constants and a violation of  the Equivalence Principle.  Effects that  are  currently searched  
in  laboratory, space-based 
experiments and astronomical observations.

The suitable constants are the   
fine-structure constant 
$\alpha = e^2/(\hbar c)$ 
and the electron-to-proton mass ratio
$\mu = m_{\rm e}/m_{\rm p}$, or different
combinations of them with the  proton gyromagnetic ratio $g_{\rm p}$  \cite{FK07}.

\begin{figure}[htb]
\centering
\includegraphics[width=2.787in]{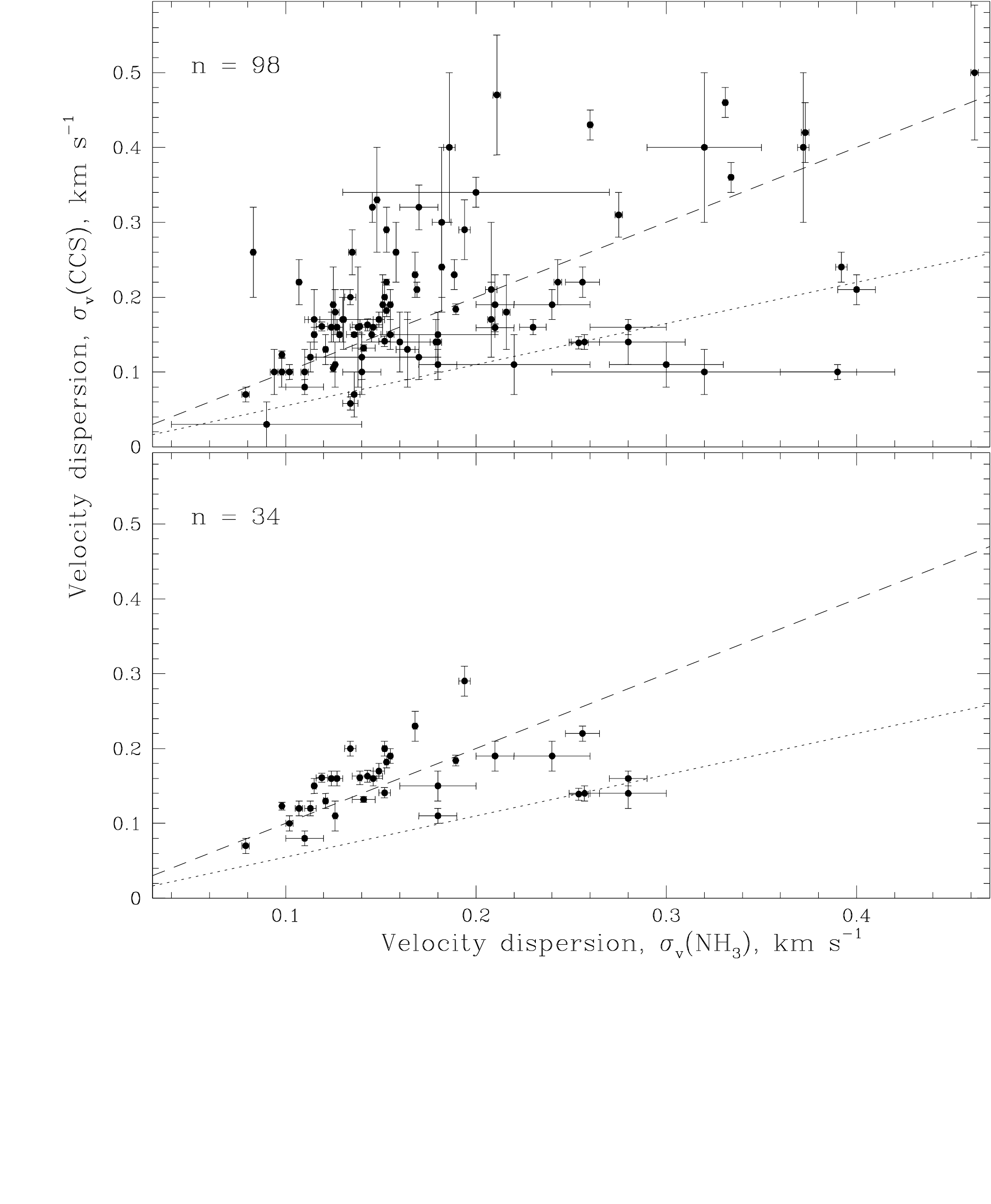}
\caption{  {\it Upper panel}: CCS ($2_1-1_0$) versus NH$_3$(1,1) linewidths
for cores in the Perseus molecular cloud from   \cite{Rosolowsky08}.
The error bars represent $1\sigma$ standard
deviations. {\it Lower panel}: selected sample with single-component
profiles. The dashed and dotted lines are 
the boundaries for  pure turbulent and pure thermal line broadening.  
\label{fg1}}
\end{figure}

\begin{figure}[htb]
\centering
\includegraphics[width=2.787in]{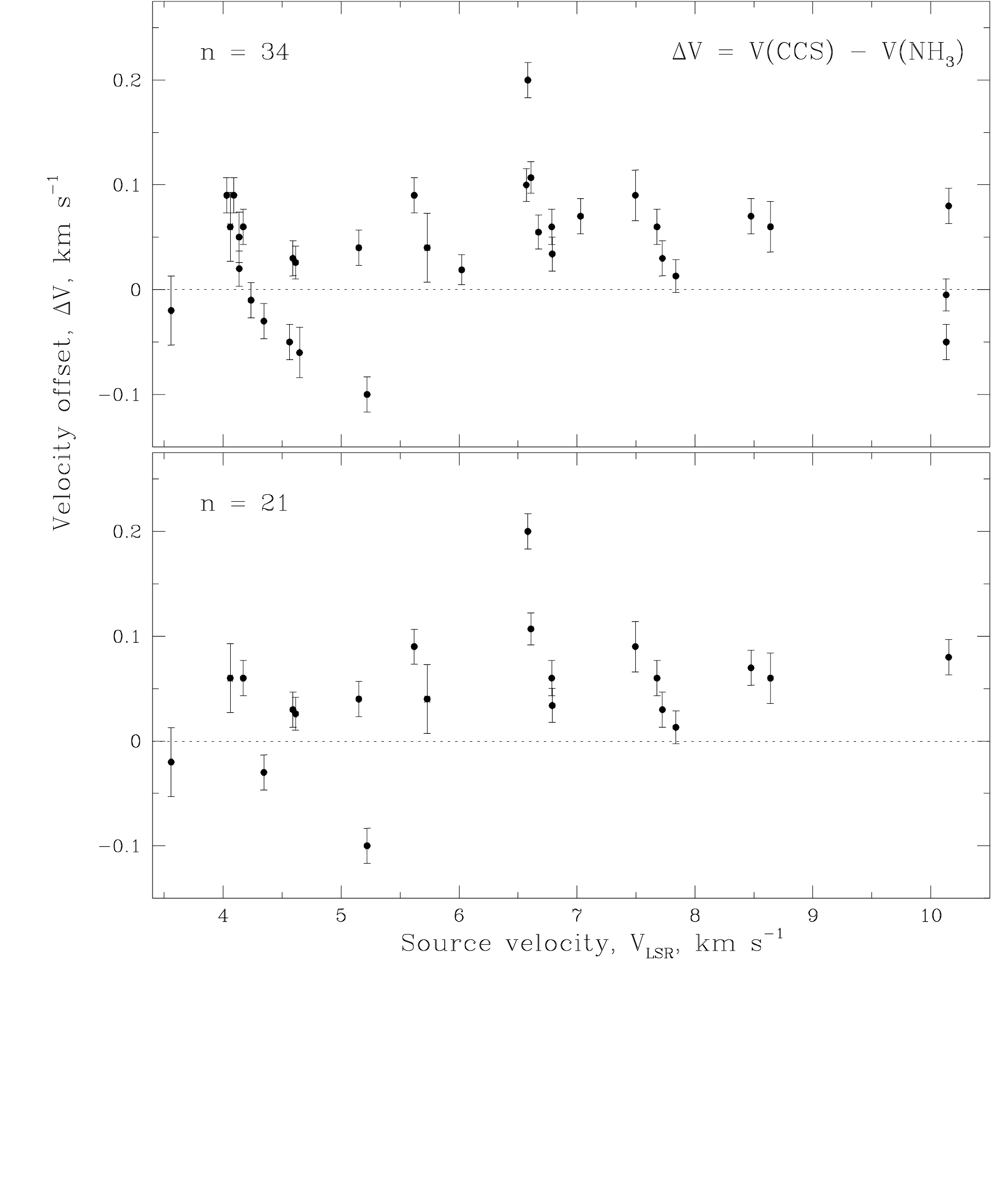}
\caption{ {\it Upper panel}: 
Velocity offset $\Delta V_{{\rm CCS-NH}_3}$ 
versus the  radial velocity 
for points shown in Fig.~\ref{fg1}.
{\it Lower panel}: Same as the upper panel but for the points with consistent broadening between the lines in Fig 1.
 The vertical error bars include both random and systematic errors caused by
the uncertainties of the adopted rest frequencies.
\label{fg2}}
\end{figure}

Accurate laboratory atomic clocks experiments constraint  the temporal variation of $\alpha$ of
$\dot{\alpha}/\alpha = (-1.6\pm2.3)\times10^{-17}$ yr$^{-1}$ 
\cite{RHS08}.
Analysis of quasar absorption-line systems provided $\Delta\alpha/\alpha = (-4.6 \pm 1.1)\times10^{-6}$ 
\cite{Murphy08} or  $|\Delta\alpha/\alpha| \le  6  \times  10^{-6}$ \cite{Srianand08,Molaro08}.
Here $\Delta \alpha/\alpha = (\alpha' - \alpha)/\alpha$, with $\alpha$
denoting the  value of the fine-structure constant in the laboratory and  $\alpha'$ the
specific absorption/emission line system of a galactic or extragalactic object,  
the same definition is applied to \dmm.  
Being linearly extrapolated to redshift $z \sim 2$, or
$t \sim 10^{10}$ yr, the laboratory bound 
  leads to 
$|\Delta\alpha/\alpha| \le  4 \times10^{-7}$ 
which is  below the astronomical limits or claims for variability. 
However, laboratory experiments and quasar absorption spectra probe
 different time scales and regions of the universe, 
and the connection between them is somewhat model dependent 
\cite{MB04}.

The direct  laboratory estimate of time variation of $\mu$  
  gives  
$\dot{\mu}/\mu = (1.6 \pm 1.7)\times10^{-15}$ yr$^{-1}$ \cite{Blatt08}. 
Astrophysical estimates of changes of $\mu$ at high redshifs 
are controversial with  claims of nonzero and zero changes:
$\Delta \mu/\mu = (-2.4\pm0.6)\times10^{-5}$ was inferred from
the analysis of the H$_2$-bearing clouds at 
$z = 2.6$ and $3.0$ \cite{RBH06}, 
but this value was not confirmed later \cite{King08}.
More stringent bounds  are  obtained at 
redshifts $z = 0.68$ and  $z = 0.89$ from the analysis of radio-frequency transitions in
NH$_3$  and other molecules such as CO, HCO$^+$,  HC$_3$N providing 
$\Delta \mu/\mu = (-0.6\pm1.9)\times10^{-6}$,  
$\Delta \mu/\mu = (0.08 \pm 0.47)\times10^{-6}$ \cite{FK07,Henkel09},
which favors a non-variability of $\mu$ at a  level of $\sim1.5\times10^{-6}$.

In these   studies it is implicitly assumed that the rate of time variations 
dominates over  possible spatial variations.
However,  if scalar fields trace 
the gravitational field inhomogeneities spatial variation could be present as well \cite{Barrow05}.  The circumstance of 
spatial variations of constants are tested by measurements of the atomic clock transition  in laboratories on Earth orbiting in  
the changing gravitational potential of the Sun. 
The laboratory experiments constrain  the couplings of $\alpha$ and $\mu$
to the gravitational field   at a level of  
$k_\alpha,_\mu \sim 10^{-6}$, 
$ 10^{-5}$ \cite{Blatt08}.
To reconcile these results with quintessence, chameleon-type models were suggested  
which allow scalar
fields to evolve on cosmological time scales today and to have 
simultaneous strong couplings of order unity to matter  \cite{Khoury04}. 
In these models  the mass of the scalar field 
depends on the local matter density, which 
 explains why cosmological scalar fields, such as quintessence,
are not detectable in local tests of the Equivalence Principle. On Earth
 in a dense
environment  the mass of the field can be sufficiently large and
$\varphi$-mediated interactions are short-ranged. On the other hand,
the cosmological matter density  is about $10^{30}$ times smaller and 
the mass of the scalar field can be very low, of order of $H_0$,  allowing  the field to evolve cosmologically.

In this paper we apply the ammonia  method for the first time to  Milky Way sources. A detailed and exaustive account has been provided in \cite{Levshakov08}.

\section{Interstellar molecules  probe  \dmm\ spatial variations}

Among  molecules,
ammonia is of particular interest due to its high
sensitivity to changes in the
electron-to-proton mass ratio, $\mu$ \cite{FK07}.
 For  NH$_3$, the sensitivity coefficient of 
the inversion transition $\nu = 23.69$ GHz 
was calculated in \cite{FK07}: 
\begin{equation}
\frac{\Delta \nu}{\nu} = 4.46\frac{\Delta \mu}{\mu}\ .
\label{S2eq2}
\end{equation}
By comparing the observed inversion frequency of NH$_3$ (1,1) with a suitable 
rotational frequency  
of another molecule arising  co-spatially, 
a limit on the spatial variation of $\mu$ can be determined:
\begin{equation}
\frac{\Delta \mu}{\mu} = 0.289\frac{V_{\rm rot} - V_{\rm inv}}{c}
\equiv 0.289\frac{\Delta V}{c}\ ,
\label{S2eq3}
\end{equation}
where $V_{\rm rot}$ and $V_{\rm inv}$ are the apparent radial velocities
of the rotational and inversion transitions, respectively.

Thus the comparison of 
the relative radial velocities of ammonia inversion lines 
and rotational transitions of another N-bearing
(N$_2$H$^+$) and C-bearing (CCS, HC$_3$N)
molecules can be used 
to set a limit on spatial variations of \dmm.
These molecules are often detected in the molecular clouds  allowing  to probe this constant  in the Milky Way. 
In molecular clouds considerable fraction of material is not participating in the process of star formation  residing in cold and dense clouds  which are well suited for accurate measurements.
These can be  low-mass     protostellar cores
containing infrared sources, or even  starless cores   without embedded luminous sources. 
The  physical conditions of the latter are of dense molecular
gas preceding gravitational collapse and when  the  gas density in the core center is 
$n_{{\scriptscriptstyle \rm H}_2} \le 10^5$ \cmm  they  can be
dynamically stable against gravitational contraction 
\cite{Keto08}.
Typical cores   have 
 mean gas density $n_{\scriptscriptstyle \rm H} \sim 
(1-2)\times10^4$ \cmm,
velocity dispersion $\sigma_v = 0.17$ \kms, kinetic temperature
$T_{\rm kin} = 11$ K, radius $R = 0.09$ pc,  mass 
${\cal M} \sim 1\ {\cal M}_\odot$  and 
NH$_3$, N$_2$H$^+$, CCS, and HC$_3$N 
are usually observed in emission.

\section{Cores in the Perseus molecular cloud}

An ammonia spectral atlas of 193 dense 
protostellar and prestellar cores of low masses in the
Perseus molecular cloud is provided by 
\cite{Rosolowsky08}. 
The spectral observations
of the cores in NH$_3$ (1,1), (2,2),
CCS ($2_1-1_0$) and CC$^{34}$S ($2_1-1_0$) lines were carried out
simultaneously using the 100-m  Green Bank Telescope (GBT).
Each target was observed in a single-pointing, frequency-switched mode.
The GBT beam size at 23 GHz is $FWHM = 31''$ or 0.04 pc at the distance of 260 pc of Perseus cloud. 
The  spectral resolution was 24 \ms and  both molecules were observed with similar angular 
and spectral resolutions.

The atlas of 193 sources is very conveniently presented on the 
website (see \cite{Rosolowsky08} for details)
  where the original spectra are shown along with
the best fitting models, model parameters and their uncertainties.
The  number of cores where both 
CCS ($2_1-1_0$) and NH$_3$ (1,1) lines were detected is 98. 
The central velocities of these lines,
being averaged with weights
inversely proportional to the variances of the measurements, reveal 
that $\Delta V_{{\rm CCS-NH}_3}$=   16 \ms.  
It was suggested \cite{Rosolowsky08} that 
this offset is due to uncertainties in the
assumed rest frequency of the CCS line which is of 13.5 \ms (cfr Table 1). 
However,  the 
unweighted averaging, which is more appropriate in the presence of unaccounted errors,  gives
${\Delta V}_{n=98} =  44\pm 13$ \ms\ 
 
 Many cores show evidence
for the multiple velocity components along the line-of-sight.   
 The selection   of  simple  profiles which   have been satisfactorily  fitted by a single 
component model provides a  sample of $n = 34$ pairs 
 The weighted mean
 for this  sample   gives ${\Delta V}^w_{n=34} =  40\pm 10$ \ms.
and
${\Delta V}_{n=34} = 39\pm 10$ \ms for the unweighted  mean \cite{Levshakov08}. 

A further refinement of the sample  could be done taking  only cores  which show line broadening values  between  pure turbulent, $\sigma_v({\rm CCS}) = \sigma_v({{\rm NH}_3})$,
and pure thermal, $\sigma_v({\rm CCS}) = 0.55 \sigma_v({{\rm NH}_3})$,
regimes, which should be the case if  both emission lines arise co-spatially.
 This requirement further reduce the  sample to  $n = 21$ clouds with
${\Delta V}^w_{n=21} =  50\pm0.013$ \ms, and
${\Delta V}_{n=21} =  48\pm 13$ \ms for the weighted and unweighted mean respectively.

The fact that scatter  remains rather constant in the various samples reflects  the  complexity of gas kinematics and
effects of chemical segregation of one molecule with respect to the
other in dense molecular cores which are the dominant sources of error.

The unweighted mean from the $n = 21$  sample could be taken 
as the best  estimate of the velocity offset for the Perseus dark cores, but in fact we could have taken any of the former samples without changing significantly the result.  When we account for  uncertainties in the
rest frequencies , we have  $\Delta V_{{\rm CCS-NH}_3}$=$48\pm 13\pm13.5$ \ms.
Other observations in the Pipe nebula \cite{Rathborne08} and in Infrared dark clouds \cite{Sakai08} are examined in \cite{Levshakov08}.  8 pairs in the Pipe Nebula provide
$\Delta V_{{\rm CCS-NH}_3}$=$53\pm11_{\rm stat}\pm13.5_{\rm sys}$ \ms.    36  NH$_3$/N$_2$H$^+$ and 
 27 NH$_3$/HC$_3$N pairs observed in  Infrared Dark Clouds, which are much more massive than the dense cores, provide
$\Delta V_{{\rm N_2H^+ -NH}_3}$$= 148\pm32_{\rm stat}\pm13.6_{\rm sys}$ \ms\ and 
$\Delta V_{{\rm HC_3N-NH}_3}$=$115\pm37_{\rm stat}\pm31_{\rm sys}$ \ms, respectively, thus supporting the results found in the Perseus region.

\section{The  L183  dense core}

Recently   Pagani et al.  \cite{Pagani09} 
have compared    the  \ndhpb and \nddpb  transitions with the transitions of NH$_3$  
in the dark cloud L183. 
The whole elongated dense core of L183   has  been fully mapped in  \ndhpb and \nddpb \juzb lines with 
the IRAM 30-m telescope  with  velocity resolution  in the range 30--50 \ms. 
Spatial resolution ranges from 33 arcsec at 77 GHz to 9 arcsec at 279 GHz. 
The spatial sampling is 12 arcsec for the main prestellar core and 15 arcsec for the southern 
and northern extension  of the prestellar core. 
Observations of NH$_3$ (1,1) and (2,2) inversion lines  have been performed  at  the Green Bank 100-m
telescope   with velocity sampling of 20
m\,s$^{-1}$,   angular resolution of $\sim$35 arcsec  and spatial sampling of 24 arcsec all over the source.

Extremely narrow molecular lines observed in this cold dark cloud
provide a sensitive spectroscopic tool 
to track small systematic velocity gradients  and  to align 
values of poorly known frequencies with more reliable ones.  For the latter,
NH$_3$ is the molecule   of choice since it is  stable  and its frequencies are accurately  measured in the laboratory.
NH$_3$ (1,1) average frequency from the whole HFS  is $\nu$~=~23\,694\,495\,487 ($\pm$48) Hz   \cite{Kukolich67},
or  $\nu$~=~23\,694\,495\,481 $\pm$ 22 Hz from the revision of 
Hougen \cite{Hougen72}. 
Thus NH$_3$ provides  a reference transition known with a precision of  few 10$^{-10}$ or  $\sim 0.6$ \ms\ .

Ammonia and diazenylium have the same chemical origin, starting from N$_2$ and are well-known 
to be coexistent as discussed by e.g. \cite{Tafalla02,Tafalla04}. 
In L183  the velocity along the dense filament is constantly changing suggesting  a flow towards the 
prestellar cores and  a rotation of the filament around its vertical axis. 
NH$_3$~(1,1), \ndhpb and \nddpb \juzb all trace exactly the same gradients and it seems 
therefore compulsory that their velocities be identical as there is no obvious possibility that 
the velocity gradients be exactly parallel but offset from each other, especially 
in the case of a  cylinder rotation. Finally, the fact that \nddpb velocity centroids 
are almost identical with those of \ndhpb indicates that the different opacities of the lines 
are not introducing any measurable bias here.
  
A similar behaviour of all the three  molecules    
shows  that these  share the same volume of the cloud and undergo the same macroscopic velocity shifts. 
 
 \begin{table}[t!]
\caption{Molecular transitions and uncertainties,
$\varepsilon_v$,  
The numbers in parentheses correspond to $1\sigma$ errors.}
\label{tbl-1}
\begin{tabular}{llr}
\noalign{\smallskip}
\multicolumn{1}{c}{Transition} & 
\multicolumn{1}{c}{$\nu_{\rm rest}$,} & 
\multicolumn{1}{c}{$\varepsilon_v$,} \\
 & \multicolumn{1}{c}{GHz} &   \multicolumn{1}{c}{\ms} \\
\noalign{\smallskip}
\hline
\noalign{\medskip}
CCS  $J_N=2_1-1_0$ & 22.344033(1)&  13.4 \\
NH$_3$ $(J,K)=(1,1)$ & 23.694495481(22)&  0.6  \\
HC$_3$N   $J=5-4$ & 45.4903102(3)& 2.8  \\
N$_2$H$^+$$J=1-0$ & 93.173777(4)&  13.5  \\
 \end{tabular}
\end{table}

On the basis that all the three species are spatially coexistent and trace the same velocities 
  Pagani et al \cite{Pagani09}  
used the observed  offset to  adjust the frequencies of \ndhpb and \nddpb to that of NH$_3$. 
For the reference position alone, the difference is of 
$\Delta V_{{\rm N_2H^+-NH}_3}$=$40.8 \pm 0.56$  m~s$^{-1}$ 
and 
$\Delta V_{{\rm N_2H^+-NH}_3}$=$40.8 \pm12.9$  ~m~s$^{-1}$  on  65 common positions .  
For the  three transitions of \nddp,  the direct comparison of the reference position with NH$_3$(1,1) spectrum yields
\juz\,: 37.7 $\pm$ 0.85 \ms
\jdu\,: 47.7 $\pm$ 0.92 \ms
\jtd\,: 63.6 $\pm$ 4.7  \ms  and similar offsets are found for \nddpb (cfr \cite{Pagani09} for details). 


 The rest frequencies for the molecule N$_2$H$^+$ in \cite{Pagani09}  analysis 
 were taken from  \cite{Caselli95} and 
are known with precision of $\sim 7$ kHz. 
However, an error of 4 KHz  is   given in 
the Cologne Database for Molecular Spectroscopy (CDMS). 
The  difference between the data from \cite{Caselli95} and the CDMS frequencies is
$-12$ \ms and the   Pagani et al velocity offset between
N$_2$H$^+$ and NH$_3$  should be corrected by this difference when adopting the CDMS rest frequency.

\section{Discussion}
\subsection{A safe bound for  \dmm\ }
At the high spectral resolution of these observations  
the errors in the molecular line position measurements 
are mainly restricted by the uncertainties
in laboratory frequencies, $\varepsilon_\nu \sim 1$ kHz, 
which correspond to the $V_{\rm LSR}$ uncertainties of 
$\varepsilon_v \sim 10$ \ms\ 
Taking into account that \dmm $\sim 0.3\Delta V/c$,
these errors implies  a sensitivity in   \dmm\  
at a level  of $\sim 10^{-8}$, i.e.  about 100 times more sensitive  
than the \dmm\ estimate deduced at $z = 0.68$ 
\cite{FK07}. This level of accuracy
can be achieved when  assuming  that 
molecules are  co-spatially
distributed within the cloud, and are observed
simultaneously with the same receiver, beam size, system temperature, and
velocity resolution.
Violation of any of these conditions leads to
shifts of the line centers.
Overall the  observations in the Perseus dense cores as well as in the L183 core  
allow  a maximum offset between  ammonia and the other molecules  of $\Delta V \le 100  $ \ms\  
in the most conservative way, which  corresponds to   \dmm $ \le 1\times 10^{-7}$,  
i.e.  an order of magnitude more sensitive than previous
astronomical constraints on $ \mu$.

This  firm bound  
is    in conflict with the value
$\Delta \mu/\mu = (-24\pm6)\times10^{-6}$ obtained from molecular
hydrogen H$_2$ absorption lines at $z \sim 3$ \cite{RBH06}.
However,   
the variability of $\mu$ was not confirmed at the level
of $|\Delta \mu/\mu| \leq 4.9\times10^{-5}$  \cite{Wendt09} and, 
recently,  a more stringent limit on \dmm\ was found 
at $z \sim 3$, 
$\Delta \mu/\mu = (2.6\pm3.0)\times10^{-6}$ \cite{King08}.

These  extragalactic measurements of \dmm\  are performed in systems in which gas densities 
are similar to those in the Milky Way
clouds. Thus also in the high-$z$ absorbers 
the dependence of $\mu$ on $\rho$ is extremely weak, and the value of \dmm\
in quasar absorbers is expected to be at the same level as
in the interstellar clouds, i.e. $\sim 10^{-7}$.  If a  \dmm\ is found above this value, 
then the variation should be temporal rather than spatial.

\subsection{Hints of  variation of \dmm\ ?}

Taken at face value the measured $\Delta V$ in Perseus and L183 being similar while involving 
different molecules show the possibility of a  positive  shifts between
the line centers of NH$_3$ and other molecules of the order of 40 \ms . 
This  would imply  a \dmm $\sim 4\times10^{-8}$.

In  theories which
consider a direct coupling between the local matter density and
the scalar fields  driving  varying constants  
the spatial change of the constants 
is proportional to the local gravitational potential
\cite{Olive08} (and references cited therein).
The effect of scalar fields on $\mu$ can be expressed as
\begin{equation}
\frac{\Delta \mu}{\mu} = k_\mu \Delta \Phi\ ,
\label{S4eq5}
\end{equation}
where $k_\mu$ is a dimensionless coupling constant of a very light 
scalar field to the local gravitational potential $\Phi$, and
$\Delta \Phi$ is the difference of the gravitational potentials
between two measurement points.
Since in our case $\Delta \Phi\ \sim 10^{-7}$, a value of \dmm = $4 \times10^{-8}$ would require  $k_\mu \sim 1$,
which   is inconsistent   with the $  k_\mu \le 10^{-5}$estimated  from  
atomic clock experiments \cite{Blatt08} .

However,  chameleon  models are specifically conceived to circumvent this type of conflict. 
The  baryon masses and coupling constants are
strongly dependent on the local matter density, $\rho$, \cite{Olive08}. 
The dynamics of the scalar field in these models depends on $\rho$:
in low-density environments it is determined by the scalar field potential
$V(\varphi)$, whereas in high-density regions like the Earth's surface
it is set by the matter-$\varphi$ coupling. The range of the scalar-mediated
force for the terrestrial matter densities is then very short, less than 1 mm
\cite{Olive08}. 
Because of the strong $\varphi$-coupling to matter 
the measurements of the frequency drifts in the
atomic clock experiments ought to be insensitive to the changes in the
gravitational potential at Earth caused by the eccentricity of Earth's orbit.
  
  We note that the difference of matter densities between 
the terrestrial environment 
($\rho_\oplus \sim 3\times10^{24}$ GeV~\cmm) and, e.g., dense molecular cores 
($\rho_{\rm cloud} \sim 3\times10^{5}$ GeV~\cmm) is of 20 orders of magnitude.
Thus,   \dmm\ variation \dmm = $4 \times10^{-8}$ require 
 models which treat the $\varphi$-mediated
force as a short-range force depending on the matter density.
In accord with model predictions, the estimated
value of the scalar field is much less than one.  In this way
the measured spatial variation of \dmm\ does
not contradict the laboratory studies on atomic clocks
due to the  extremely different density environment
in the terrestrial measurements and in the interstellar medium.

\section{Conclusions}

\begin{enumerate}
\item  Several molecules and in different environments in the Milky Way are used to probe \dmm  
variations by means of the NH$_3$ method.   This is  done by using CCS  measured in almost  
one hundred of dense Perseus clouds and  \ndhpb and \nddpb 
in a detailed high spatial resolution survey  of the dense core  L183.  

\item These  observations provide evidence that NH$_3$ does not change by more than   
100 \ms\ allowing us to place a robust bound of  \dmm $\le10^{-7}$  which is more than  
an order of magnitude  more stringent than the extragalactic bounds of the same quantity.

\item Both sets of comparison point for a common offset  of  about 40 \ms. 
Although it is possible that this  offset  results
from rest  frequency uncertainties the fact that several 
transitions are involved suggest that the effect might be real.    
An intrinsic   shift of 40 \ms   would correspond to  \dmm $\sim 4\times10^{-8}$.  

\item To cope with solar system experiments the effect  would require    chameleon-type scalar
field models which predict a strong dependence of masses and coupling
constants on the ambient matter density.  
  
\item New observations to verify this  possibility  are solicited,  and some are underway at 
radio telescopes  of Medicina, Effelsberg and Nobeyama.

\end{enumerate}

{\bf Acknowledgments}
The authors are grateful to Eric Rosolowsky, Takeshi
Sakai, and Jill Rathborne who sent us additional comments
on their observations and data reduction. We
thank Paola Caselli, Irina Agafonova, Keith Olive, Alexander Lapinov and
Thomas Dent for useful comments and discussions.
SAL is supported by the RFBR grant 09-02-00352-a. 


 \end{document}